\documentclass[useAMS,referee,usenatbib]{paper}

\usepackage{epsfig,graphicx,bbm,mathtools,multirow}
\usepackage[shortlabels]{enumitem}
\usepackage{booktabs}
\usepackage{color}

\newcommand{\bO}{\boldsymbol{O}}

\newcommand{\bd}{\boldsymbol{d}}

\newcommand{\be}{\boldsymbol{e}}

\newcommand{\bX}{\boldsymbol{X}}

\newcommand{\bz}{\boldsymbol{z}}

\newcommand{\bzero}{\boldsymbol{0}}

\newcommand{\bxi}{\boldsymbol{\xi}}

\newcommand{\btheta}{\boldsymbol{\theta}}

\newcommand{\bgamma}{\boldsymbol{\gamma}}

\newcommand{\bzeta}{\boldsymbol{\zeta}}

\newcommand{\balpha}{\boldsymbol{\alpha}}

\newcommand{\bepsilon}{\boldsymbol{\epsilon}}

\newcommand{\bx}{\boldsymbol{x}}

\usepackage[figuresright]{rotating}

\title[Latent Class Analysis for Time-to-event Data]{Latent Class Analysis with Semi-parametric Proportional Hazards Submodel for Time-to-event Data}

\author
{Teng Fei$^{1}$, John Hanfelt$^2$ and Limin Peng$^{2, *}$\email{lpeng@emory.edu} \\
$^1$Department of Epidemiology and Biostatistics, Memorial Sloan Kettering Cancer Center,\\ 485 Lexington Avenue, New York, New York, 10017, U.S.A. 
\\
$^2$Department of Biostatistics and Bioinformatics, Emory University,\\ 1518 Clifton Road Northeast, Atlanta, Georgia 30322, U.S.A.}

\begin{document}

\volume{}
\pubyear{}
\artmonth{}
\doi{}

\label{firstpage}


\begin{abstract}
Latent class analysis (LCA) is a useful tool to investigate the heterogeneity of a disease population with time-to-event data. We propose a new method based on non-parametric maximum likelihood estimator (NPMLE), which facilitates theoretically validated inference procedure for covariate effects and cumulative hazard functions. We assess the proposed method via extensive simulation studies and demonstrate improved predictive performance over standard Cox regression model. We further illustrate the practical utility of the proposed method through an application to a mild cognitive impairment (MCI) cohort dataset. 

\end{abstract}

%


\maketitle

\section{Introduction}

The nature of heterogeneity has been recognized in a number of diseases, such as mild cognitive impairment (MCI) and prostate cancer. Usually, the clinicians classify the patients into certain disease \textit{subtypes}, such as the amnestic and the non-amnestic subtypes of MCI \citep{winblad04}, where each subtype represents a particular etiology with potentially unique pattern of disease progression. It is of critical clinical interest to understand the heterogeneity of disease population and its implications for the onset of clinical events, which will contribute to better prediction of time-to-event, such as predicting the time to dementia for MCI patients based on baseline patient characteristics. 

Common practice to address this interest is regression analysis, such as fitting a Cox proportion hazard model \citep{cox72}. Typically, traditional survival models attempt to explain the survival distribution for the whole population by a single model with unified covariate effects, such that different values or levels of covariates indicate earlier or later onset of disease. For a heterogeneous population, however, traditional survival models are oversimplified. This is because the heterogeneity of disease population indicates different underlying etiologies, which means the disease progression and the importance of associated risk factors can vary among disease subtypes. In statistical modeling, this implies varied baseline hazard functions and covariate effects for different disease subtypes. 

Latent class analysis (LCA) 
is a useful tool to address the above challenges in analyzing survival data of heterogeneous populations. Extended from finite mixture models \citep{mclachlan00}, the LCA framework is able to incorporate class-specific survival submodels to capture heterogeneous patterns in disease progression, and a class membership probability submodel which addresses the uncertainty of belonging to certain latent subtypes given patient characteristics. In addition, the latent classes (or subtypes) defined by LCA are jointly determined by the membership probability submodel and the class-specific survival submodels, which means the obtained latent classes are data-driven with high relevance to the survival outcome of interest. 

{Various mixture models have been proposed in the past few decades for the clustering analysis of survival data. The two-component mixture cure models \citep[for example]{kuk1992,mclachlan1994,lambert2010} were studied to investigate differences in survival between cured and uncured population. However, mixture cure models assume only two classes in the population, which is not applicable if there are more than two classes.
In addition, mixture Weibull models \citep[for example]{buvcar2004,mair2009} and mixture exponential models \citep[for example]{hilton2018} were proposed to investigate heterogeneous lifetime distribution for two or more underlying classes. Nevertheless, these methods are less flexible due to the imposed parametric assumption of survival distribution. }

{In recent development of mixture modeling for biomedical data, large efforts were made in the development of joint latent class models of survival data and other phenotypes, such as longitudinal data \citep[for example]{lin2002,PL09,PL17} and responses to a questionnaire \citep{larsen04}. Typically, these models assume conditional independence between phenotypes and survival data given the class membership, which ignores within-class correlation between survival time and other phenotypes. Under this assumption, these methods may obtain redundant latent classes which are largely attributed to the heterogeneity of phenotypes, instead of the time-to-event of interest.
Hypothetically, some phenotypes can have high heterogeneity in disease population but little correlation with time-to-event. Under such hypothetical setting, the joint models will still recognize the latent classes with respect to the phenotypes, which can hardly contribute to our research question of interest in understanding the heterogeneity in survival. In terms of survival submodels, the majority of existing joint latent class models use class-specific Cox proportional hazard model. \citet{PL17} utilized Weibull distribution, piecewise constant with limited number of jumps, and cubic M-splines to formulate baseline hazard functions, which creates challenges in model specification. In contrast, \citet{lin2002} and \citet{larsen04} incorporated unspecified baseline hazard functions which imposes weaker assumptions. However, both methods assumed common covariate effects on survival for different latent classes, which is less flexible in capturing the heterogeneity of covariate effects. In addition, little attention was paid to deriving asymptotic theories of the estimators for the semi-parametric latent class models.}

{Motivated by the limitations of the existing methods, we provide a semi-parametric framework for the latent class analysis of survival data, to investigate the heterogeneity of disease population and its implications for disease progression. We impose weaker assumptions of unspecified baseline cumulative hazard function, which improves flexibility compared to the existing methods with parametric assumptions. We also enables class-specific covariate effects in both latent class probability submodel and survival submodel, such that the heterogeneity can be better detected and interpreted. In addition, our framework focuses on the heterogeneity in time-to-event distribution, such that the resulting latent class patterns are not interfered by other phenotypes with limited contribution to understanding survival heterogeneity.}

{Technically, we utilize non-parametric maximum likelihood estimator \citep[NPMLE]{zeng07} approach to incorporate the the infinite-dimensional component of baseline cumulative hazard function. Due to the finite mixture structure of latent class problems, the finite-dimensional and infinite-dimensional components are entangled in the likelihood function, which creates further challenges of establishing asymptotic properties and conducting variance estimation. We address this difficulty using an approach similar to that used by \citet{mao2017}. To handle unobservable latent class labels, we derive a stable expectation-maximization (EM) algorithm which can be easily assembled by existing software or algorithm. According to our numerical experience, the algorithm is robust to initialization and achieves impressive results with non-informative initial values. Moreover, asymptotic theories are rigorously established by empirical process arguments \citep{van96} and semi-parametric efficiency results \citep{bickel1993}. We also provide alternative strategies for inference, based on either information matrix, or profile likelihood \citep{murphy00}. Furthermore, we give recommendations in model selection criteria in selecting the most appropriate number of latent classes.}

\section{Data, notation and models}
\label{sec:datamodel}

\subsection{Data and notations}

Let $T$ and $C$ respectively denote time to event of interest and time to independent censoring of $T$. Let $\bx$ denote a $p \times 1$ vector of baseline covariates. 
Define $\tilde{T} = T \wedge C$ and $\Delta = I(T \le C)$. The observed data consist of $n$ independent and identically distributed replicates of $\bO = (\tilde{T},\Delta,\bx)$, denoted by $\{\bO_i = (\tilde{T}_i,\Delta_i,\bx_i),i = 1,\ldots,n\}$. The latent classes are denoted by an unobservable $L \times 1$ vector of binary indicators, $\bxi = (\xi_{1},\ldots,\xi_{L})$, where $L$ is the number of latent classes, and $\xi_l = 1$ if belonging to the $l$th class and 0 otherwise. 


\subsection{The assumed models}
In this work we assume that the marginal density of time-to-event observation $(\tilde T, \Delta)$ can be captured by a finite mixture model \citep{mclachlan00} with $L$ components
\[
    f(\tilde T, \Delta) = \sum_{l=1}^L p_l f_l(\tilde T, \Delta),
\]
where $p_l$ is the probability of belonging to class $l$ and $f_l(\tilde T,\Delta)$ is the class-specific density of $(\tilde T, \Delta)$ for class $l$. Our modeling strategy involves a class membership probability submodel for $p_l$ and a class-specific survival submodel for $f_l(\tilde T, \Delta)$. 

For the class membership probability submodel, we utilize a standard latent polytomous logistic regression model \citep{BR97} to account for the effect of baseline covariates on the relative frequency of latent classes:
\begin{equation}
\rmn{Pr}(\xi_{l}=1|\bx) = p_l(\bx;\balpha) = \frac{\exp(\tilde\bx^T\balpha_l)}{\sum_{d=1}^{L}\exp(\tilde\bx^T\balpha_d)}, l = 1,\ldots,L,
\label{eq:margmod}
\end{equation}
where $\tilde \bx = (1,\bx^T)^T$, $\balpha_1 = \bzero$ for identifiability consideration, and $\balpha = (\balpha_2,\ldots,\balpha_L)^T$ is the vector of unknown parameters with length $(p+1) \times (L-1)$.

For the class-specific survival submodel, we propose a semi-parametric class-specific proportional hazards model. Without loss of generality, we let the first class be the reference class with hazard function
    $\lambda(t|\xi_{1} = 1) = \lambda_{0}(t)\exp(\bar{\bx}^T\bzeta_{1}),$
where $\lambda_0(t)$ is the unspecified baseline hazard function for the reference class, $\bar{\bx}$ is a $q \times 1$ subvector of $\bx$ with $q \le p$, and $\bzeta_1$ is the $q \times 1$ unknown covariate effect in the reference class model. For other classes $l = 2,\ldots,L$ that is not the first class, we assume
    $\lambda(t|\xi_{l} = 1) = \lambda_{0}(t)\exp\{a_l + \bar{\bx}^T(\bzeta_{1}+\bzeta_{l})\}$, 
where $\exp(a_l)$ is a constant ratio between the baseline hazard functions of class $l$ and class 1, and $\bzeta_l$ is the $q \times 1$ difference of covariate effects between class $l$ and class 1. Let $\bz_{l} = (\bar{\bx}^T,\bzero_{(q+1)\times(L-1)}^T)^T\cdot {I(l = 1)} + (\bar{\bx}^T,(\be_{l-1}\otimes \tilde{\bx})^T)^T \cdot {I(l > 1)}$ and $\bgamma = (\bzeta_{1}^T, a_2, \bzeta_2^T, a_3, \bzeta_3^T,\ldots,a_L, \bzeta_L^T)^T$, where $\bzero_d$ represents a $d-$vector of zeros, $\be_{l-1}$ represents a $(L-1)-$vector whose $(l-1)$th element is 1 and other elements are zero, ${\tilde{\bx} = (1,\bar{\bx}^T)^T}$, and $\otimes$ denotes Kronecker product operator. Then it follows a universal expression of the class-specific hazard functions for class 1 to class $L$
\begin{equation}
    \lambda(t|\xi_{l} = 1) = \lambda_{0}(t)\exp(\bz_l^T\bgamma), \ l = 1,\ldots,L,
    \label{eq:survmod}
\end{equation}
where $\bgamma$ is the vector of unknown parameters with length $q \times L + (L-1)$. 
The according class-specific density of $(\tilde{T},\Delta)$ 
satisfies
    $f_l(\tilde{T}, \Delta|\bx;\btheta) = \{\lambda_{0}(\tilde{T})\exp({\bz}_{l}^T\bgamma)\}^{\Delta} \exp\{-\Lambda_{0}(\tilde{T})\exp({\bz}_{l}^T\bgamma)\},$
where $\Lambda_0(t) = \int_0^t\lambda_0(s)ds$ and $\btheta = \{\bgamma^T,\Lambda(\cdot)\}^T$. {Then the finite mixture framework implies that the conditional density of $(\tilde T, \Delta)$ given $\bx$ satisfies}
\begin{equation}
    f(\tilde T, \Delta | \bx; \balpha, \btheta) = \sum_{l=1}^L p_l(\bx,\balpha)f_l(\tilde T,\Delta | \bx;\btheta).
    \label{eq:finitemix}
\end{equation}

\section{Estimation and inference}
\label{sec:estimation}
Based on the assumed models in Section \ref{sec:datamodel}, in this section we derive the likelihood function and the associated estimation and inference procedures. {Due to the complications caused by the missingness of latent class memberships $\bxi$, and the non-parametric assumption of the baseline cumulative hazard function $\Lambda_0(\cdot)$, it is not straightforward to maximize the likelihood function for the observed data. As a natural solution, we utilize non-parametric maximum likelihood estimators (NPMLE) technique to account for the unobservable $\bxi$ by an Expectation-Maximization (EM) algorithm, and to facilitate inference for the non-parametric estimator $\hat{\Lambda}(\cdot)$ of $\Lambda_0(\cdot)$.}

\subsection{Observed data likelihood}

Under the assumed finite mixture model (\ref{eq:finitemix}), and submodels (\ref{eq:margmod}) and (\ref{eq:survmod}) we obtain the observed data likelihood
\begin{equation}
        L(\balpha,\bgamma,\Lambda;\bO) = \prod_{i=1}^n \bigg\{\sum_{l=1}^L p_l(\bx_i;\balpha) \{\lambda(\tilde{T}_i)\exp({\bz}_{il}^T\bgamma)\}^{\Delta_i} \exp\{-\Lambda(\tilde{T}_i)\exp({\bz}_{il}^T\bgamma)\} \bigg\}f_{\bX}(\bx_i),
     \label{eq:obslik}
\end{equation}
where $f_{\bX}(\cdot)$ is the density function of $\bx$. Note that $f_{\bX}(\bx_i)$ is a constant with respect to unknown parameters $\balpha,\bgamma$ and $\Lambda$, thus is omitted in further derivations.

\subsection{EM algorithm for point estimation}

Assuming $\bxi$ are observed, the complete data likelihood corresponding to the observed data likelihood (\ref{eq:obslik}) satisfies
\[
        L_c(\balpha,\bgamma,\Lambda;\bxi,\bO) = \prod_{i=1}^n \prod_{l=1}^L\bigg\{ p_l(\bx_i;\balpha) \{\lambda(\tilde{T}_i)\exp({\bz}_{il}^T\bgamma)\}^{\Delta_i} \exp\{-\Lambda(\tilde{T}_i)\exp({\bz}_{il}^T\bgamma)\} \bigg\}^{I(\xi_{il}=1)}.
\]
We further treat $\Lambda(\cdot)$ as piecewise constant between observed event times. That is, $\Lambda(t) = \sum_{j:t_j \le t}\Lambda\{t_j\}$ with $\Lambda\{t_j\} = d_j$, where ${t_1 < t_2 < \ldots < t_m}$ are {distinct} uncensored event times. {Denote the cumulative hazard function $\Lambda(t_j)$, at $t_j, j = 1,\ldots,m$, as $\Lambda_j$.} Then the corresponding log complete data likelihood satisfies
\begin{equation}
\begin{split}
    \ell_c(\balpha,\bgamma,\Lambda;\bxi,\bO) =& \sum_{j=1}^m \sum_{l=1}^L\xi_{(j)l} \bigg\{ \log \Lambda\{t_j\} + \bz_{(j)l}^T\bgamma - e^{\bz_{(j)l}^T\bgamma}\Lambda_j \bigg\}\\
    &- \sum_{j=1}^m \sum_{k:t_j \le \tilde T_k < t_{j+1}} I(\Delta_k = 0)\sum_{l=1}^L\xi_{kl}e^{\bz_{kl}^T\bgamma}\Lambda_j\\
     &+\sum_{i=1}^n\sum_{l=1}^L\xi_{il}\log p_l(\bx_i;\balpha),
\end{split}
\label{eq:cloglik}
\end{equation}
where $\xi_{(j)l}$ and $\bz_{(j)l}$ represents the membership indicator $\xi_l$ and covariate vector $\bz_l$ for the observation with uncensored failure time $t_j, j = 1,\ldots,m$. 

In the E-step, {we calculate the expectation, $E\{\ell_c(\balpha,\bgamma,\Lambda;\bxi,\bO) | 
\bO,\balpha^{(j)},\bgamma^{(j)},\Lambda^{(j)}\}$, of the log complete data likelihood \eqref{eq:cloglik}, conditioned on observable data $\bO$ and the current estimates of unknown parameters $\balpha^{(j)},\bgamma^{(j)},\Lambda^{(j)}$ at the arbitrary $j$th iteration. Because of the simplicity of \eqref{eq:cloglik} with respect to $\bxi$, it is straightforward to see $E\{\ell_c(\balpha,\bgamma,\Lambda;\bxi,\bO) | 
\bO,\balpha^{(j)},\bgamma^{(j)},\Lambda^{(j)}\} = \ell_c\{\balpha,\bgamma,\Lambda;\hat E(\bxi),\bO\}$, where $\hat E(\xi_{il}) \equiv E(\xi_{il} | \bO_i;\balpha^{(j)},\bgamma^{(j)},\Lambda^{(j)})$.}
Note that $E(\xi_{il} | \bO_i;\balpha,\bgamma,\Lambda) = \rmn{Pr}(\xi_{il} = 1|\bO_i;\balpha,\bgamma,\Lambda)$ is the posterior membership probability which can be derived by {Bayes' Rule} 
$\rmn{Pr}(\xi_{il} = 1|\bO_i;\balpha,\bgamma,\Lambda) = {\rmn{Pr}(\xi_{il} = 1, \tilde T_i, \Delta_i |\bx_i;\balpha,\bgamma,\Lambda)}/$
${\rmn{Pr}(\tilde T_i, \Delta_i |\bx_i;\balpha,\bgamma,\Lambda)}.$ 
That is, 
\begin{equation}
    \hat{E}(\xi_{il}) = \rmn{Pr}(\xi_{il} = 1|\bO_i;\balpha^{(h)},\bgamma^{(h)},\Lambda^{(h)}) = \frac{p_l(\bx_i;\balpha^{(h)})f_l(\tilde{T}_i, \Delta_i|\bx_i;\bgamma^{(h)},\Lambda^{(h)})}{\sum_{d=1}^Lp_d(\bx_i;\balpha^{(h)})f_d(\tilde{T}_i, \Delta_i|\bx_i;\bgamma^{(h)},\Lambda^{(h)})}.
    \label{eq:hatE}
\end{equation}
The resulting conditional expectation $ \ell_c\{\balpha,\bgamma,\Lambda;\hat E(\bxi),\bO\}$, denoted by $Q(\balpha,\bgamma,\Lambda)$, serves as the target function to be maximized in the subsequent M-step.

In the M-step, we adopt a profile likelihood strategy to maximize $Q(\balpha,\bgamma,\Lambda)$ by profiling out $\Lambda$, where $\Lambda$ is treated as an $m$-dimensional unknown parameter $\Lambda\{t_k\} = d_k, k=1,\ldots,m.$ 
First, with fixed $\balpha$ and $\bgamma$, we find $\hat \Lambda(t;\bgamma) = \arg\max_{\Lambda}Q(\balpha,\bgamma,\Lambda)$ by solving
\[
     \frac{\partial}{\partial d_k} Q(\balpha,\bgamma,\Lambda) = \frac{1}{d_k} - \sum_{i:\tilde T_i \ge t_k}\sum_{l=1}^L\hat E(\xi_{il})e^{\bz_{il}^T\bgamma} = 0, \quad k = 1,\ldots,m.
\]
That is, $\hat d_k(\bgamma) = \{\sum_{i:\tilde T_i \ge t_k}\sum_{l=1}^L\hat E(\xi_{il})e^{\bz_{il}^T\bgamma}\}^{-1}, k = 1,\ldots,m$ and 
\begin{equation}
    \hat \Lambda(t;\bgamma) = \sum_{k:t_k \le t}\hat d_k(\bgamma) = \int_0^t \frac{\sum_{i=1}^n dN_i(s)}{\sum_{i=1}^n\sum_{l=1}^L \hat E(\xi_{il})Y_i(s)e^{\bz_{il}^T\bgamma}},
    \label{eq:breslow}
\end{equation}
where $N(t) = I(\tilde T \le t, \Delta = 1)$ and $Y(t) = I(\tilde T \ge t).$
Then by plugging in $\hat \Lambda(t;\bgamma)$, we obtain the profile complete data log likelihood $Q_p(\balpha,\bgamma) \equiv Q\{\balpha,\bgamma,\hat\Lambda(t;\bgamma)\}$:
\begin{equation}
\begin{split}
     Q_p(\balpha,\bgamma)
     =& \sum_{i=1}^n \sum_{l=1}^L \int_0^{t^*}\hat E(\xi_{il}) \bigg\{ \log \frac{1}{\sum_{i=1}^n\sum_{l=1}^L \hat E(\xi_{il})Y_i( s)e^{\bz_{il}^T\bgamma}} + \bz_{il}^T\bgamma\bigg\}dN_i(s)  \\
     &+\sum_{i=1}^n\sum_{l=1}^L\hat E(\xi_{il})\log p_l(\bx_i;\balpha),
\end{split}
\label{eq:pcloglik}
\end{equation}
where $t^*$ is a finite constant satisfying $t^* > t_m.$ Then it is straightforward to find $\hat \balpha = \arg\max_{\balpha} Q_p(\balpha,\bgamma)$ and $\hat \bgamma = \arg\max_{\bgamma} Q_p(\balpha,\bgamma)$ by solving 
\[
    \frac{\partial}{\partial \balpha} Q_p(\balpha,\bgamma)= \sum_{i=1}^n\sum_{l=1}^L\hat E(\xi_{il}) \frac{\partial}{\partial \balpha}\log p_l(\bx_i;\balpha) = \bzero
\]
and 
\[
    \frac{\partial}{\partial \bgamma} Q_p(\balpha,\bgamma) = \sum_{i=1}^n\sum_{l=1}^L \int_0^{t^*}\hat E(\xi_{il})\bigg(\bz_{il} - \frac{\sum_{j=1}^n\sum_{k=1}^L\hat E(\xi_{jk})Y_{j}(u)\bz_{jk}\exp(\bz_{jk}^T\bgamma)}{\sum_{j=1}^n\sum_{k=1}^L\hat E(\xi_{jk})Y_{j}(u)\exp(\bz_{jk}^T\bgamma)} \bigg)dN_i(u) = \bzero.
\]
{It is straightforward to show that solving $\frac{\partial}{\partial \balpha} Q_p(\balpha,\bgamma) = 0$ reduces to fitting a weighted multinomial logistic regression with weights $\hat{E}(\bxi)$, which can be easily implemented by R package \texttt{VGAM} \citep{yee10}. In addition, equation $ \frac{\partial}{\partial \bgamma} Q_p(\balpha,\bgamma) = 0$ is equivalent to a weighted partial score equation for the proportional hazard model with weights $\hat{E}(\bxi)$. We choose not to use existing Cox regression software to solve $ \frac{\partial}{\partial \bgamma} Q_p(\balpha,\bgamma) = 0$, which would automatically account for the pseudo ties caused by repeatedly counting each observed event (indexed by $i$) for multiple latent classes (indexed by $l$), making the resulting estimates not accurately based on the estimating equation $ \frac{\partial}{\partial \bgamma} Q_p(\balpha,\bgamma) = 0$. Instead, we implement an efficient Newton-Raphson algorithm under \texttt{Rcpp} environment \citep{eddelbuettel2011} to ensure that the estimator is a rigorous solution of $ \frac{\partial}{\partial \bgamma} Q_p(\balpha,\bgamma) = 0$.}

We initialize the EM algorithm with an initial guess of $\hat{E}(\bxi)$, which can be obtained from random guess or informative ways such as K-means clustering of $\tilde T$.
Then we repeat the M-step and E-step until the stopping criterion is satisfied. We propose to use an Aitken acceleration-based stopping criterion as described in \citet[page 52]{mclachlan00}. Denote $l^{(k)}$ as the logarithm of the observed-data likelihood (\ref{eq:obslik}) evaluated using the parameter estimation at the $k$th iteration. Define
        $a^{(k)} = (l^{(k+1)} - l^{(k)})/(l^{(k)} - l^{(k-1)})$
    and
        $l_A^{(k+1)} = l^{(k)} + (l^{(k+1)}-l^{(k)})/(1-a^{(k)})$.
The algorithm is stopped when $|l_A^{(k+1)} - l_A^{(k)}| < tol$, where $tol$ is the tolerance parameter. In practice, we let $tol = 10^{-7}$ to ensure convergence to a local optimum. 

\subsection{Asymptotic properties and variance estimation}

In this section, we establish the consistency and asymptotic normality using NPMLE arguments similar to those used in \citet{zeng2006} and \citet{mao2017}. First we give the following regularity conditions:\\
(C1) There exists $t^* > 0$ such that $\rmn{Pr}(C = t^*) > 0$ and $\rmn{Pr}(C > t^*) = 0$;\\
(C2) For $l = 1, \ldots, L$, $\rmn{Pr}(\xi_l = 1|\bx;\balpha) \in (0,1)$.\\
(C3) $||\balpha_0|| < \infty$; $||\bgamma_0|| < \infty$;  $||z_l|| < \infty$ for $l = 1,\ldots,L$; $\Lambda_0$ is continuously differentiable with $\Lambda'(t) > 0$ on $[0,t^*]$,
where $||\cdot||$ denotes the Euclidean norm.

\noindent Conditions (C1)-(C3) are reasonable in practical applications. Condition (C1) is commonly satisfied by administrative censoring, which also helps prove the uniform consistency of $\Lambda(\cdot)$ on $[0,t^*]$. Condition (C2) ensures that the latent class membership probabilities $p_l(\bx;\balpha)$ is greater than zero, which further guarantees that $\log p_l(\bx;\balpha)$ has a finite lower bound. Condition (C3) assumes the smoothness of $\Lambda(\cdot)$ and the boundedness of $\balpha_0$, $\bgamma_0$ and baseline covariates $\bx$. Proofs of the following two theorems are provided in Web Appendix A. 
\begin{theorem}
Under regularity conditions (C1)-(C3), $\hat\balpha$, $\hat\bgamma$ and $\hat\Lambda(\cdot)$ are strongly consistent. That is, 
$||\hat\balpha - \balpha_0|| + ||\hat\bgamma - \bgamma_0|| + \sup_{t \in [0,t^*]}|\hat\Lambda(t) - \Lambda_0(t)| \to 0$
almost surely.
\label{thm:consistency}
\end{theorem}


\begin{theorem}
Under regularity conditions (C1)-(C3), { $\sqrt{n}(\hat\balpha-\balpha_0)$ and $\sqrt{n}(\hat\bgamma-\bgamma_0)$ converges to multivariate zero-mean Gaussian distributions; $\sqrt{n}\{\hat\Lambda(t)-\Lambda_0(t)\}$ converges to a univariate zero-mean Gaussian process on $t \in [0,t^*]$. In addition, $\hat{\balpha}$ and $\hat{\bgamma}$ are semiparametric efficient as defined in \citet{bickel1993}.}
\label{thm:weakconv}
\end{theorem}
{Variance estimation can be conducted} based on the information matrix of the observed-data profile log-likelihood \citep{murphy00}, defined by $\rmn{pl}(\balpha,\bgamma) \equiv \ell\{\balpha,\bgamma,\hat{\Lambda}(\balpha,\bgamma);\bO\},$ where $\hat{\Lambda}(\balpha,\bgamma) = \rmn{argmax}_\Lambda \ell(\balpha,\bgamma,\Lambda;\bO)$. Given the point estimates $(\hat\balpha^T,\hat\bgamma^T)^T$, we obtain $\hat{\Lambda}(\hat\balpha,\hat\bgamma)$ by running the aforementioned EM algorithm with $\hat\balpha$ and $\hat\bgamma$ fixed, and only updating $\hat\Lambda(\cdot)$ {by formula \eqref{eq:breslow}} {and $\hat{E}$ by formula \eqref{eq:hatE}} until convergence. Then it follows an estimation of the profile log-likelihood $\hat{\rmn{pl}}(\hat\balpha,\hat\bgamma) = \ell\{\hat\balpha,\hat\bgamma,\hat{\Lambda}(\hat\balpha,\hat\bgamma);\bO\}.$ Let $\hat{\rmn{pl}}_j(\hat\balpha,\hat\bgamma)$ be the subject $j$'s contribution to $\hat{\rmn{pl}}(\hat\balpha,\hat\bgamma)$. The covariance matrix of $\hat\btheta = (\hat\balpha^T,\hat\bgamma^T)^T \in \mathbbm{R}^{r}$, where $r = (p+1)\times(L-1)+q\times L + L -1$, can be estimated by the inverse of
\[
\sum_{j=1}^n
 \left(\begin{array}{c} \frac{\hat{\rmn{pl}}_j(\hat\btheta + h_n\bepsilon_1) - \hat{\rmn{pl}}_j(\hat\btheta - h_n\bepsilon_1)}{2h_n}\\
    \vdots \\
    \frac{\hat{\rmn{pl}}_j(\hat\btheta + h_n\bepsilon_{r}) - \hat{\rmn{pl}}_j(\hat\btheta - h_n\bepsilon_{r})}{2h_n}  \end{array}\right)^{\otimes 2},
\]
where $\bepsilon_k$ is the $k$th canonical vector in $\mathbbm{R}^r$, $\bd^{\otimes 2} = \bd \bd^T$, and $h_n$ is a constant of order $n^{-1/2}$. In the numerical studies, we used $h_n = 5n^{-1/2}$ as used by \citet{gao19}. Unlike the numerical approximation of Hessian matrix as used in \citet{murphy00}, we utilize the outer product of the first order numerical differences, which is computationally more affordable and guarantees that the resulting covariance matrix estimator is positive definite. {Alternatively, an analytical consistent variance estimator can be constructed based on similar arguments as in \citet{zeng2006}, which allows inference for $\hat{\Lambda}(t)$ in addition to $\hat{\balpha}$ and $\hat{\bgamma}$. Details about the analytical variance estimator are provided in Web Appendix A. Compared to the numerical variance estimator based on profile likelihood, the analytical variance estimator typically requires inverse matrix computation for a covariance matrix with much higher dimension due to the inclusion of cumulative hazard function, which might cause less stable numerical performance. Thus, we report inference for $\hat{\balpha}$ and $\hat{\bgamma}$ based on the profile likelihood approach in our simulation and real application analysis. In contrast, inference for $\hat{\Lambda}(t)$ is based on the analytical approach.} 


\subsection{Selecting the number of latent classes}

In practice, it is usually of interest to determine the number of latent classes, $L$, using data-driven criteria. Standard model selection criteria for likelihood-based latent class methods include the Akaike information criterion (AIC) and the Bayesian information criterion (BIC). It is also common to use entropy-based criteria, such as integrated complete-data likelihood \citep[ICL-BIC]{biernacki00} and classification entropy extended BIC \citep[CE-BIC]{hart20}. A standardized entropy index \citep{muthen02}, defined as
\[
    1-\frac{\sum_{i=1}^n\sum_{l=1}^L \hat{E}(\xi_{il}|\bO_i;\hat\balpha,\hat\bgamma,\hat\Lambda) \{-\log \hat{E}(\xi_{il}|\bO_i;\hat\balpha,\hat\bgamma,\hat\Lambda) \}  }{n\log L},
\]
is another commonly used metric to assess the level of uncertainty of latent classes in a fitted model. When the latent classes are well separated, the estimated posterior class membership probability $\hat{E}(\xi_{il}|\bO_i;\hat{\balpha},\hat{\bgamma},\hat{\Lambda})$ is close to either one or zero, such that the corresponding standardized entropy index is close to one. According to our simulation analysis detailed in Section \ref{sec:modelselect}, BIC is the most effective criterion to determine $L$ for the proposed method.


%

\subsection{{Assessing the prediction performance}}
\label{sec:prediction}
{Define $S(t|\bx,\xi_l=1) = \rmn{Pr}(T \ge t | \bx, \xi_l=1), l = 1,\ldots,L$ as the class-specific survival function, and $G(u) = \rmn{Pr}(C \ge u)$ as the survival function of the censoring at time $u$. We evaluate the prediction performance of the proposed latent class model by the Brier Score, defined as $E[\{I(T \ge t) - {\hat{S}(t|\bx)}\}^2]$, where
\begin{equation}
    \hat{S}(t|\bx) = \sum_{l=1}^L \hat{\rmn{Pr}}(\xi_l=1|\bx)\hat{S}(t|\bx,\xi_l=1) = \sum_{l=1}^Lp_l(\bx;\hat{\balpha}) \exp\{-\hat{\Lambda}(t)\exp(\bz_l^T\hat{\bgamma})\}
    \label{eq:predicted}
\end{equation}
is the predicted survival probability at time $t$ given baseline covariates $\bx$. Here the predicted survival probability $\hat{S}(t|\bx)$ can be interpreted as a weighted summation of predicted class-specific survival probabilities $\hat{S}(t|\bx,\xi_l=1) = \exp\{-\hat{\Lambda}(t)\exp(\bz_l^T\hat{\bgamma})\}$, with estimated class membership probabilities $\hat{\rmn{Pr}}(\xi_l=1|\bx) = p_l(\bx;\hat{\balpha})$ as weights. In practice, we observe $Y(t) = I(\tilde{T} \ge t)$ instead of $I(T \ge t)$. To account for the censoring status of $\tilde{T}$, we adapt the two types of estimators of the Brier Score as defined by formulae 12 and 13 in \citet{PL14}, namely data-based Brier Score 
\[
    \hat{\rmn{BS}}_1(t) = \frac{1}{n}\sum_{i=1}^n \bigg\{ \frac{I(\tilde{T}_i > t)}{\hat{G}(t)}\{1 - \hat{S}(t|\bx_i)\}^2 + \frac{\Delta_i I(\tilde{T}_i \le t)}{\hat{G}(\tilde{T}_i)}\{0 - \hat{S}(t|\bx_i)\}^2 \bigg\}
\]
and model-based Brier Score 
\begin{equation*}
    \begin{split}
        \hat{\rmn{BS}}_2(t) =& \frac{1}{n}\sum_{i=1}^n \bigg[ I(\tilde{T}_i > t)\{1 - \hat{S}(t|\bx_i)\}^2 + \Delta_i I(\tilde{T}_i \le t)\{0 - \hat{S}(t|\bx_i)\}^2 \\
        & + (1-\Delta_i) I(\tilde{T}_i \le t)\bigg\{\{1 - \hat{S}(t|\bx_i)\}^2\frac{\hat{S}(t|\bx_i)}{\hat{S}(\tilde{T}_i|\bx_i)} + \{0 - \hat{S}(t|\bx_i)\}^2\bigg(1-\frac{\hat{S}(t|\bx_i)}{\hat{S}(\tilde{T}_i|\bx_i)}\bigg)\bigg\}\bigg].
    \end{split}
\end{equation*}
Here an estimate $\hat{G}(\cdot)$ of the survival function for censoring can be obtained by either Kaplan-Meier or regression models.}

{In numerical analysis, we conduct 5-fold cross validation, fit models on the training set, and estimate the Brier Score $\hat{\rmn{BS}}_j^{(f)}(t), j = 1,2, f=1,\ldots,5$ for the testing set of the $f$th cross-validation fold for a given range of $t$. Then we report the average Brier score $\overline{\hat{\rmn{BS}}}_j(t) = \frac{1}{5}\sum_{f=1}^5 \hat{\rmn{BS}}_j^{(f)}(t)$ among folds to assess the prediction performances. We use Kaplan-Meier estimator to estimate $\hat{G}(\cdot)$ in our estimation of Brier Scores.}

\section{Simulation study}
\label{sec:simulation}
We conducted simulation studies to evaluate the finite-sample performance of the proposed method in terms of parameter estimation, and selecting the number of classes $L$. In addition, we compared the proposed method and the standard proportional hazard model in terms of goodness-of-fit and prediction. 
With $L = 2$ or 3, we generated a two-dimensional baseline covariate vector $\bx = (x_1, x_2)$, where $x_1$ is a binary $Bernoulli(0.5)$ random variable and $x_2$ is a continuous $Uniform(0,1)$ random variable. Then the latent class label vector $\bxi$ was generated from a $Multinomial(1,\{p_1(\bx;\balpha),\ldots,p_L(\bx;\balpha)\}^T)$ distribution following model (\ref{eq:margmod}). Given latent classes, the time-to-event $T$ was generated from class-specific distribution function $F_T(t|\xi_l = 1) = 1-\exp\{0.1(1-e^{t})\exp(\bz_l^T \bgamma)\}, l = 1,\ldots,L$ derived from model (\ref{eq:survmod}) with $\lambda_0(t) = 0.1(e^{t}-1)$. Then we generated independent censoring time $C$ as the minimum of an $Exponential(r)$ variable and a $Uniform(5,6)$ variable. 

Table \ref{tab:simuconfig} summarizes the choice of $r$, $\balpha$ and $\bgamma$ in five simulation scenarios. 
For scenarios with $L=2$ (I,II,III,IV), scenario (I) served as a benchmark with relatively light censoring rate ($r=0.1$) and less overlapped survival distributions ($a_2 = 2$) among the two classes. In contrast, scenario (II) created more overlapped survival distributions ($a_2 = 0$) while scenario (III) created heavy censoring ($r=0.6$). Scenario (IV) considered a special situation where covariate $x_1$ had a large effect size ($\alpha_{2,1} = -4$) on class probability $p_l(\bx;\balpha)$ but zero covariate effect ($\zeta_{1,1} = \zeta_{2,1} = 0$) in survival submodel, while $x_2$ had zero covariate effect ($\alpha_{2,2} = 0$) on class probability but a large effect size ($\zeta_{1,2} = -3, \zeta_{2,2} = 6$) in survival submodel. In our description later, the scenario (IV) is refer to as the scenario with ``separation of covariate effects in submodels''. Compared to scenario (I), scenario (IV) had slightly heavier censoring with similar overlapped level of survival distributions among the two classes.  With three latent classes, scenario (V) was comparable to scenario (I) in terms of censoring and the overlapping among class-specific survival distributions. Empirical metrics of censoring and overlapping among classes for the five scenarios can be found in Table \ref{tab:simudemo}.

\begin{table}
    \caption{Choices of parameters in the five simulation scenarios.} 
    \centering
    
    \begin{tabular}{l l c c c c c c c c}
    \toprule
         & & Censoring & \multicolumn{2}{c}{Parameters in  } &  \multicolumn{5}{c}{ }\\
         & & parameter & \multicolumn{2}{c}{model (\ref{eq:margmod}) $\balpha$} & \multicolumn{5}{c}{Parameters in model (\ref{eq:survmod}) $\bgamma$ }\\
        \cmidrule(lr){3-3} \cmidrule(lr){4-5} \cmidrule(lr){6-10} 
        \multicolumn{2}{l}{Simulation scenarios} & $r$ & $\balpha_2$ & $\balpha_3$ &  $\bzeta_{1}$ & $a_2$ & $\bzeta_{2}$ & $a_3$ & $\bzeta_{3,1}$\\
        \midrule
        \multirow{4}{*}{$L = 2$} & scenario (I) & 0.1 & ($\log(2)$,0,0) & \multirow{4}{*}{NA} & (-2,0) & 2 & (2,2) & \multirow{4}{*}{NA} & \multirow{4}{*}{NA}\\
        & scenario (II) & 0.1 & ($\log(2)$,0,0) & & (-2,0) & 0 & (2,2) & & \\
        & scenario (III) & 0.6 & ($\log(2)$,0,0) & & (-2,0) & 2 & (2,2) & & \\
        & scenario (IV) & 0.1 & (2,-4,0) & & (0,-3) & 0.5 & (0,6) & & \\
        \midrule
        $L=3$& scenario (V) & 0.1 & (0,-0.5,0) & (0,0,0.5) & (-2,-2) & 2 & (2,2) & 4 & (4,4)\\
        \bottomrule
    \end{tabular}

    \label{tab:simuconfig}
\end{table}


\subsection{Estimation of parameters}

To evaluate parameter estimation, we conducted 10000 simulations, with sample size $n=1000$ for scenarios (I)-(IV) and sample sizes $n=$ 1000, 2000 and 3000 for scenario (V).
To initialize the algorithm, we used a perturbed $\hat{E}(\bxi)$ from the true latent class labels $\bxi$. In addition, the variance estimation for $\{\hat\balpha^T,\hat\bgamma^T\}^T$ was conducted using the profile likelihood approach, while the variance estimation for $\hat{\Lambda}(\cdot)$ was conducted using the observed-data log-likelihood approach.
We seldom observed non-convergent estimates defined as the outlying point estimates whose $L_2$ norms $\sqrt{|\hat\balpha - \balpha_0|^2 + |\hat\bgamma - \bgamma_0|^2}$ were greater than the median $L_2$ norm out of 10000 results plus 5 times median absolute deviation (MAD). Table \ref{tab:simudemo} displays convergence rate, median standardized entropy index, and median censoring rate for different simulation scenarios out of 10000 simulations. Table \ref{tab:simudemo} indicates that compared to the benchmark scenario (I), more mixed survival distributions (II), heavier censoring (III), or larger number of $L$ (V) would result in more non-convergent results. In addition, heavier censoring (III) would also result in a lower standardized entropy index, suggesting that censoring intensified the fuzziness of the mixtures. {Moreover, scenario (IV) displays similar level of mixture as scenario (I), with a slightly higher censoring rate.}

\begin{table}
    \caption{Convergence rate, median standardized entropy index and median censoring rate out of 10000 simulations for the five simulation scenarios.}
    \centering
    
    \begin{tabular}{l l c c c c}
    \toprule
        \multicolumn{2}{l}{Simulation scenarios} & Sample size & Convergence & Median entropy & Median censoring\\
        \multirow{4}{*}{$L = 2$} & scenario (I) & 1000 & 97.66\% & 0.7686 & 11\% \\
        & scenario (II) & 1000 & 97.34\% & 0.4348 & 17\%\\
        & scenario (III) & 1000 & 96.07\% & 0.6220 & 38\%\\
        & scenario (IV) & 1000 & 97.40\% & 0.7771 & 19\%\\
        \midrule
        & & 1000 & 98.65\% & 0.7602 & 15\% \\
        $L=3$ & scenario (V) & 2000 & 96.74\% & 0.7838 & 15\% \\
        &  & 3000 & 94.06\% & 0.7785 & 15\% \\
        \bottomrule
    \end{tabular}

    \label{tab:simudemo}
\end{table}

The simulation results for four representative parameters, $\alpha_{2,2}$, $\zeta_{1,1}$, $a_2$ and $\Lambda(3)$, are shown in Table \ref{tab:simuresult}. Full results for all unknown parameters are available in supplementary Tables S.1 and S.2. {As observed, under scenario (I) and (IV) the proposed estimator achieved very small median biases and accurately estimated standard errors. The coverage probabilities of the 95\% confidence intervals are close to 0.95 for both regression coefficient $\hat{\balpha}$, $\hat{\bgamma}$ and infinite-dimensional $\hat{\Lambda}(t)$. Compared to scenario (I) and (IV), fuzzier mixture pattern in scenario (II) and heavier censoring in scenario (III) result in larger median biases for most parameters. In addition, a slight underestimation of the standard errors is observed for scenario (II) and scenario (III), such that the coverage probabilities are slightly lower than 0.95, in particular for $\hat{a}_2$. For simulation (V) with three latent classes, the estimation tends to be unstable with smaller sample size 1000, showing higher biases in the proportionality parameters $\hat{a}_2$ and $\hat{a}_3$ and regression parameters $\hat{\zeta}_{21}$ and $\hat{\zeta}_{22}$. This is probably due to insufficient sample size, which in particular damages estimation for the parameters corresponding to the second class which overlaps with both class 1 and class 3. As sample size grows to 2000 and 3000, an improvement in median biases and coverage probabilities is observed. However, compared to scenarios (I)-(IV) with two classes, the proposed method requires a larger average sample size from each class to detect the mixture pattern of time-to-event distribution.}



\begin{table}
	 \caption{Median bias (M.Bias), standard deviation (SE), median standard error estimate (SEE), and coverage probability (CP) of parameters 	$\hat{\alpha}_{2,2}$, $\hat{\zeta}_{1,1}$, $\hat{a}_2$ and $\hat{\Lambda}(3)$ out of 10000 simulations.}
    \centering
    
    \begin{tabular}{l c c c c c c c c c}
    \toprule
         $n$ & Scenarios & \multicolumn{4}{c}{$\hat{\alpha}_{2,2}$} & \multicolumn{4}{c}{$\hat{\zeta}_{1,1}$}\\
          \cmidrule(lr){3-6}\cmidrule(lr){7-10}
          & & M.Bias & SE & SEE & CP &  Bias & SE & SEE & CP \\    
         1000 & (I) & -0.020 & 0.302 & 0.296 & 0.949 & -0.024 & 0.199 & 0.201 & 0.956 \\
         1000 & (II) & -0.007 & 0.522 & 0.500 & 0.945 & -0.045 & 0.318 & 0.314 & 0.958\\
         1000 & (III) & -0.037 & 0.410 & 0.380 & 0.936 & -0.063 & 0.427 & 0.403 & 0.963\\
         1000 & (IV) & 0.010 & 0.665 & 0.696 & 0.968 & -0.011 & 0.205 & 0.207 & 0.952\\
         1000 & (V) & 0.039 & 0.581 & 0.515 & 0.908 & -0.042 & 0.236 & 0.216 & 0.938\\      
         2000 & (V) & 0.034 & 0.389 & 0.371 & 0.929 & -0.022 & 0.151 & 0.146 & 0.946\\
         3000 & (V) & 0.036 & 0.311 & 0.305 & 0.940 & -0.014 & 0.121 & 0.118 & 0.946 \\
         & & \multicolumn{4}{c}{$\hat{a}_{2}$} & \multicolumn{4}{c}{$\hat\Lambda(3)$}\\
          \cmidrule(lr){3-6}\cmidrule(lr){7-10}
         & & M.Bias & SE & SEE & CP &  Bias & SE & SEE & CP \\
         1000 & (I) & 0.011 & 0.449 & 0.412 & 0.940 & -0.010 & 0.351 & 0.344 & 0.951\\
         1000 & (II) & 0.032 & 0.451 & 0.406 & 0.926 & 0.003 & 0.545 & 0.504 & 0.949\\
         1000 & (III) & 0.016 & 0.733 & 0.616 & 0.914 & -0.018 & 0.759 & 0.661 & 0.942\\
         1000 & (IV) & 0.002 & 0.310 & 0.309 & 0.954 & 0.016 & 0.481 & 0.449 & 0.945\\
         1000 & (V) & -0.256 & 1.160 & 0.787 & 0.791 & 0.117 & 0.585 & 0.520 & 0.910\\
         2000 & (V) & -0.122 & 0.828 & 0.628 & 0.872 & 0.062 & 0.390 & 0.367 & 0.925\\
         3000 & (V) & -0.074 & 0.631 & 0.534 & 0.932 & 0.038 & 0.315 & 0.298 & 0.927\\
    \bottomrule
    \end{tabular}

    \label{tab:simuresult}
\end{table}

\subsection{Determining the number of latent classes}
\label{sec:modelselect}
We further conducted 1000 simulations for each of the five simulation scenarios with sample size $n=1000$. In each simulation, we fitted the proposed latent class model for $L \in \{2,3,4,5\}$ with algorithms initialized by K-means clustering. Then we compared model selection criteria for the models with different choices of $L$. 

\begin{figure}
    \centering
    \includegraphics[width=\textwidth]{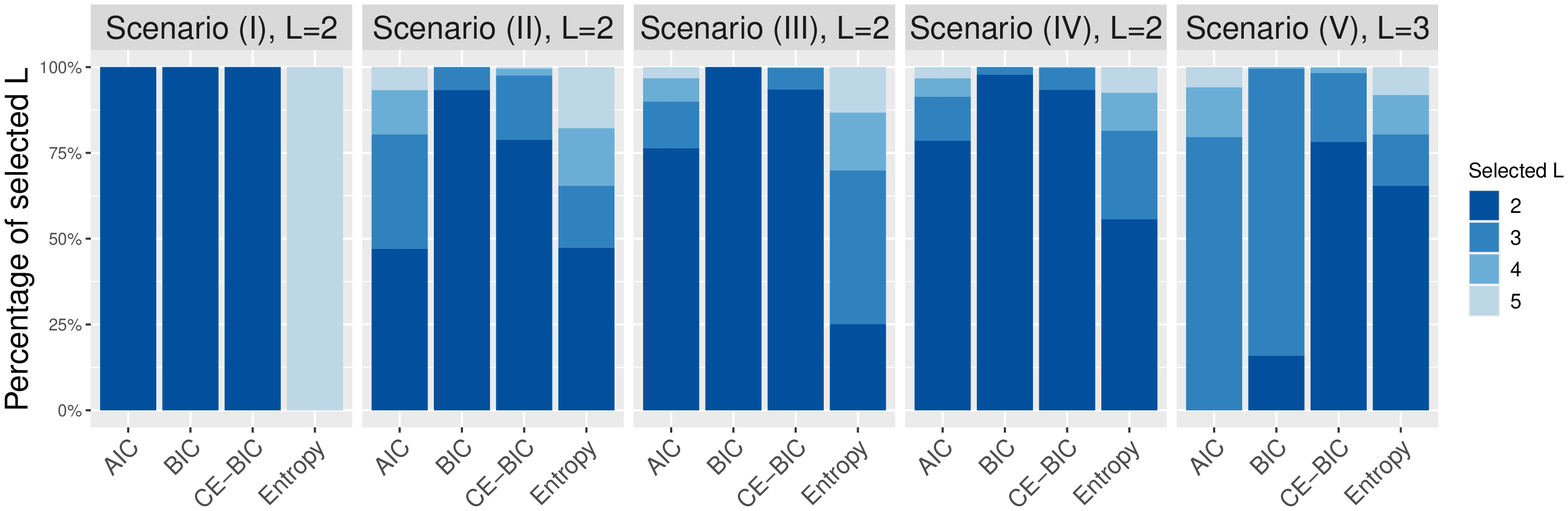}
    \caption{Percentage of latent classes selected by different model selection criteria out of 1000 simulations under simulation scenarios (I)-(V).}
    \label{fig:simu_crit}
\end{figure}


As shown in Figure \ref{fig:simu_crit}, BIC correctly selected $L$ in all 1000 simulations when the two latent classes are well separated (I), even if heavily censored (III). BIC also performed well under heavy mixture (II), with separated covariate effects in submodels (IV), and three-class (V) scenarios. Compared to BIC, AIC tended to select a larger number of latent classes, particularly for the heavy mixture scenario (II).
In terms of entropy-based criteria, we found that the standardized entropy index tended to select incorrect $L$, which also explained that the classification entropy extended BIC (CE-BIC) performed worse than the standalone BIC. Similar results were also observed when there were three latent classes in scenario (V).
The superiority of BIC over entropy-based criteria can be explained by the fact that the proposed method is a likelihood-based method. According to the performance in the five scenarios, BIC is the most effective criterion in selecting $L$. We also utilized BIC to select $L$ in our real data application in Section \ref{sec:real}.

\subsection{Goodness-of-fit and prediction}

{For each of the five simulation scenarios, we further simulated 1000 datasets with sample size 1000. For each simulated dataset, we conducted five-fold cross-validation as described in Section \ref{sec:prediction} to obtain the averaged estimates $\overline{\hat{\rmn{BS}}}_1(t)$ and $\overline{\hat{\rmn{BS}}}_2(t)$ of the Brier Score for a standard Cox regression model and the proposed latent class model. We set the upper bound of time interval $t^* = 5$ for scenarios (I) - (IV) and $t^* = 5.75$ for scenarios (V) to cover the support of time-to-event. Note that the Cox regression model is a special case of the latent class model with $L = 1$. Therefore, under the Cox regression model we have $\rmn{Pr}(\xi_1 = 1|\bx) = 1$ and the predicted survival function $\hat{S}(t|\bx) = \hat{S}(t|\bx,\xi_l=1) = \exp\{-\hat{\Lambda}(t)\exp(\bx^T\hat{\bzeta}_1)\}$ is solely based on the single class (or class 1) considered in the model.
}

\begin{figure}
    \centering
    \includegraphics[width=0.9\textwidth]{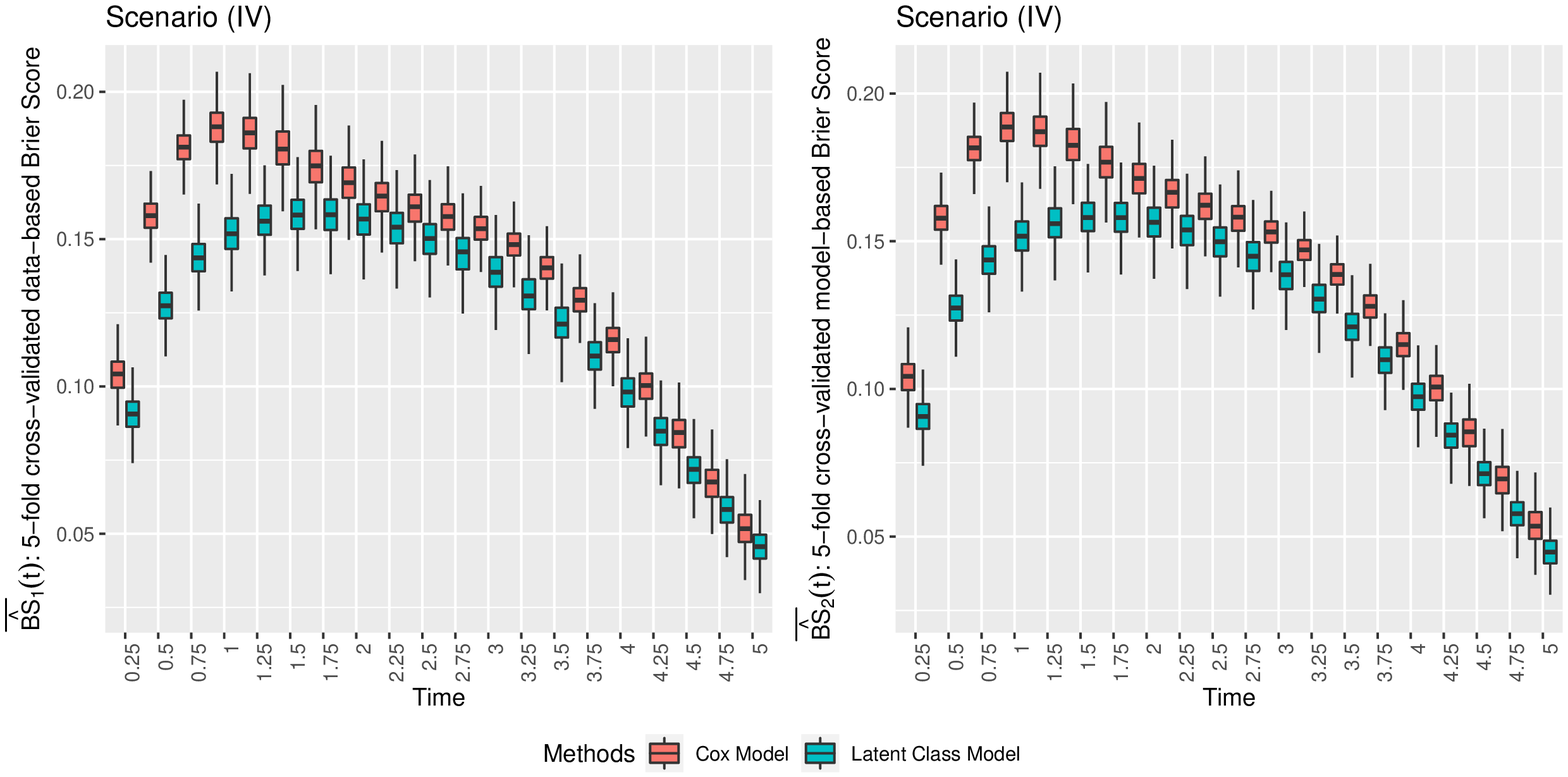}
    \caption{Boxplots for average cross-validated Brier Score $\overline{\hat{\rmn{BS}}}_1(t)$ and $\overline{\hat{\rmn{BS}}}_2(t)$, $t \in (0,5]$, from 1000 simulations under scenario (IV) with sample size 1000, for the Cox model and the proposed latent class model with $L=2$.}
    \label{fig:simu_brier}
\end{figure}

{Figure \ref{fig:simu_brier} and Supplementary Figures S.1 - S.4 shows the obtained time-dependent Brier Score estimates for scenarios (IV), (I), (II), (III) and (V), respectively. Overall, the proposed latent class model achieved consistently lower median average corss-validated Brier Score estimates than those obtained by the Cox model in all simulation scenarios. As shown in Supplementary Figures S.1 - S.4, however, only minor improvements can be recognized for scenarios (I), (II), (III) and (V), where baseline covariate effects are present in both class probability submodel and class-specific survival submodel. In contrast, the improvement is obvious under scenario (IV), where we have separation of covariate effects in submodels. Under this situation, covariates have different effects towards class membership probability and class-specific survival, which is difficult to be captured by a single-class standard Cox model.}



\section{Real data example}
\label{sec:real}
We applied our method to investigate the heterogeneity of mild cognitive impairment (MCI) using time-to-dementia data collected for 5348 patients in the Uniform Data Set between September 2005 and June 2015 by the U.S. National Alzheimer's Coordinating Center. 1501 patients developed dementia during the follow up, showing a high censoring rate of 72\%. We incorporated patients' baseline cognitive characteristics as covariates, including overall cognition (Mini-mental state examination, MMSE), executive functions (Trail making test B , TB, and Digit symbol, DS), memory (logical memory delayed, LMD, and category fluency, CF), language (Boston naming, BN), and attention (Trail making test A, TA, and digit span forward, DSF). In addition, patients' baseline number of impaired instrumental activities of daily living (IADLs), number of neuropsychiatric symptoms (NPI-Q), binary measure of depression (GDS), indicator of cerebrovascular disease (EH), and baseline age (AGE) were also included as baseline covariates. Detailed descriptions about the dataset and covariates were reported in \citet{hanfelt18}.

{High heterogeneity of the MCI population indicates that there exist MCI subgroups associated with a specific cognitive domain or domains. Thus, it is expected that the progression to dementia for different MCI subgroups are driven by their corresponding domain factors. We applied the proposed latent class model to investigate such heterogeneity in terms of the importance and effect sizes of baseline covariates.}

{We first decided the best number of classes $L$. Specifically, we fitted the proposed models with $L$ classes with random initialization for multiple times, then selected the model with the smallest BIC as the best $L$-class model. We conducted the above procedure for $L \in \{2,3,4\}$. The 2-class model obtains the smallest BIC (24481) compared to the 3-class model (24625) and the 4-class model (24797), where the BIC shows an increasing trend as $L$ increases from 2 to 4. Thus, we regard the 2-class model as the best latent class model.}

\subsection{{Summary statistics of the obtained two latent classes}}

{According to the fitted latent class model with two latent classes, we assign patients to the two classes by modal assignment. That is, we assign each patient to the class with the highest posterior membership probability $\hat{E}(\bxi)$. As Table \ref{tab:realsummary} shows, 69\% of the patients are assigned to class 1, while 31\% of the patients are assigned to class 2. Comparing the two classes, the first class had significantly smaller MMSE compared to the second class, showing better overall cognitive status. Moreover, class 1 was generally better than class 2 in most of the domain-specific scores, apart from the Boston Naming test associated with the language domain. In addition, patients in class 2 were older than those in class 1. In terms of time-to-event, patients in class 1 generally took longer than patients in class 2 to reach dementia during the follow-up, where only 18\% of patients developed dementia in class 1 but half of patients developed dementia in class 2.}



\begin{table}[]
\caption{Summary statistics of the baseline covariates for the two latent classes, based on modal assignment of class identity.}
    \centering
 \begin{tabular}{lccc}

\toprule

Covariates & Class 1, N = 3714$^1$ & Class 2, N = 1634$^1$ & p-value$^2$\\
\hline
$\tilde{T}$ & 1.83 (0.00, 3.42) & 1.08 (0.00, 2.08) & $<$0.001\\
$\Delta^3$ & 683 (18\%) & 818 (50\%) & $<$0.001\\
MMSE & -0.99 (-2.20, 0.00) & -2.09 (-3.78, -0.85) & $<$0.001\\
TB$^4$ & 0.42 (-0.22, 1.42) & 1.71 (0.52, 4.02) & $<$0.001\\
DS & -0.52 (-1.19, 0.11) & -1.38 (-2.01, -0.80) & $<$0.001\\
LMD & -1.23 (-2.05, -0.41) & -1.52 (-2.34, -0.65) & $<$0.001\\
CF & -0.75 (-1.35, -0.12) & -1.31 (-1.90, -0.73) & $<$0.001\\
BN & -0.61 (-1.88, 0.22) & -0.47 (-1.55, 0.29) & $<$0.001\\
TA$^4$ & 0.12 (-0.44, 0.90) & 0.70 (-0.07, 1.73) & $<$0.001\\
DSF & -0.29 (-0.88, 0.49) & -0.44 (-0.98, 0.39) & $<$0.001\\
EH & 224 (6.0\%) & 104 (6.4\%) & 0.6\\
IADLs & 1 (0, 2) & 4 (2, 6) & $<$0.001\\
NPI-Q & 1 (0, 2) & 2 (1, 4) & $<$0.001\\
GDS & 694 (19\%) & 279 (17\%) & 0.2\\
AGE & -0.20 (-0.81, 0.41) & 0.22 (-0.38, 0.77) & $<$0.001\\
\bottomrule
&&&\\
\multicolumn{4}{l}{$^1$ Median (IQR); n (\%)}\\
\multicolumn{4}{l}{$^2$ Wilcoxon rank sum test; Pearson's Chi-squared test}\\
\multicolumn{4}{l}{$^3$ Number of patients diagnosed with dementia}\\
\multicolumn{4}{l}{$^4$ Larger Trails B and Trails A scores indicate worse conditions.}\\
\\
\end{tabular}
\label{tab:realsummary}
\end{table}

\subsection{Parameter estimation and interpretation}

In order to demonstrate the utility of the proposed method {in investigating the heterogeneity in covariate importance and effect sizes}, we compare the point estimation, confidence interval and interpretations of the standard single-class Cox model and the proposed latent class model with $L=2$ by Table \ref{tab:realpointest}. From the Cox model ($\hat{\bzeta}$ in Table \ref{tab:realpointest}), it is clear that patients with worse baseline conditions in different cognitive domains (executive function, memory, language and attention), functional abilities, behavioral scales and aging tended to have increased hazard, or earlier onset, of dementia. {However, this overall picture revealed by the Cox regression model cannot conduct more detailed investigations on the correspondence between the MCI subtypes and the associated domain factors.}

\begin{sidewaystable}
\caption{Point estimates and 95\% confidence intervals for the covariate effects obtained by Cox model and the latent class model with two classes.}
\begin{tabular}{l l c l c l c l c l}
\toprule
& & \multicolumn{2}{c}{Cox model} & \multicolumn{6}{c}{Latent class model (2 classes)}\\
& & \multicolumn{2}{c}{(1 class)} & \multicolumn{2}{c}{Class probability} & \multicolumn{4}{c}{Class-specific survival submodel}\\
\cmidrule(lr){3-4}  \cmidrule(lr){5-6}  \cmidrule(lr){7-10}
Domains & Covariates  & $\hat{\bzeta}$ & 95\% CI & $\hat{\balpha}$ & 95\% CI & $\hat{\bzeta}_1$ & 95\% CI & $\hat{\bzeta}_1+\hat{\bzeta}_2$ & 95\% CI\\
&&&&&&&&&\\
& Intercept & NA & NA & -2.94$^{*}$ & (-5.00,-0.88) & NA & NA & 2.03$^{*}$ & (1.26,2.80)\\
\hline
Overall cognition & MMSE & -0.12$^{*}$ & (-0.15,-0.10) & -0.17 & (-0.40, 0.07) & -0.14$^{*}$ & (-0.21,-0.07) & -0.09$^{*}$ & (-0.16,-0.03)\\
\hline
\multirow{2}{*}{Executive functions} & TB & 0.08$^{*}$ & (0.05,0.12) & 0.20$^{*}$ & ( 0.00, 0.40) & -0.01 & (-0.14, 0.11) & 0.12$^{*}$ & (0.04,0.19)\\
& DS & -0.11$^{*}$ & (-0.17,-0.05) & -0.77$^{*}$ & (-1.28,-0.26) & 0.03 & (-0.20, 0.26) & -0.17$^{*}$ & (-0.30,-0.03)\\
\hline
\multirow{2}{*}{Memory} & LMD & -0.41$^{*}$ & (-0.46,-0.35) & 0.23 & (-0.40, 0.86) & -0.63$^{*}$ & (-0.80,-0.46) & -0.27$^{*}$ & (-0.39,-0.15)\\
& CF & -0.21$^{*}$ & (-0.27,-0.14) & -0.74 & (-1.73, 0.26) & -0.17 & (-0.36, 0.02) & -0.13 & (-0.29,0.03)\\
\hline
Language & BN & -0.03$^{*}$ & (-0.06,0.00) & 0.35$^{*}$ & ( 0.15, 0.55) & -0.17$^{*}$ & (-0.26,-0.07) & 0.04 & (-0.07,0.15)\\
\hline
\multirow{2}{*}{Attention} & TA & -0.04$^{*}$ & (-0.08,0.00) & -0.15 & (-0.49, 0.19) & -0.10 & (-0.21, 0.01) & 0.01 & (-0.07,0.10)\\
& DSF & 0.05 & (-0.00,0.10) & 0.06 & (-0.32, 0.44) & 0.01 & (-0.11, 0.13) & 0.07 & (-0.10,0.25)\\
\hline
Cerebrovascular disease & EH & -0.02 & (-0.23,0.18) & -1.10 & (-2.53, 0.34) & 0.39 & (-0.16, 0.93) & -0.11 & (-0.55,0.33)\\
\hline
Functional abilities & IADLs & 0.12$^{*}$ & (0.10,0.14) & 0.40$^{*}$ & ( 0.03, 0.76) & 0.21$^{*}$ & ( 0.14, 0.28) & 0.03 & (-0.05,0.10)\\
\hline
\multirow{2}{*}{Behavioral assessment} & NPI-Q & 0.06$^{*}$ & (0.04,0.09) & 0.19 & (-0.17, 0.55) & 0.11$^{*}$ & ( 0.03, 0.19) & 0.00 & (-0.08,0.08)\\
& GDS & 0.07 & (-0.07,0.21) & -0.70 & (-2.03, 0.62) & 0.09 & (-0.33, 0.50) & 0.12 & (-0.30,0.55)\\
\hline
Aging & AGE & 0.27$^{*}$ & (0.20,0.33) & 0.90$^{*}$ & ( 0.39, 1.42) & 0.38$^{*}$ & ( 0.20, 0.56) & 0.01 & (-0.20,0.22)\\
\bottomrule
&&&&&&&&&\\
\multicolumn{10}{l}{*Statistically signficant covariate effect based on 95\% confidence interval.}\\
\multicolumn{10}{l}{Higher scores on TB and TA indicated worse conditions.}\\
\end{tabular}
\label{tab:realpointest}
\end{sidewaystable}

{In contrast}, our proposed latent class model were able to capture the heterogeneous associations between baseline characteristics and dementia, with sensible clinical interpretations. According to the point estimates for the class membership probability submodel ($\hat{\balpha}$ in Table \ref{tab:realpointest}), younger MCI patients with more severe problems in language domain (BN) were more likely to belong to the first latent class, while older MCI patients with worse executive functions (TB and DS) and impaired functional abilities (IADLs) were more likely to belong to the second class.

The class-specific survival submodel revealed further heterogeneity of covariate effects ($\hat{\bzeta}_1$ and $\hat{\bzeta}_2$ in Table \ref{tab:realpointest}) on survival. First of all, we found for both classes worse baseline overall cognition (MMSE) had statistically significant effect in increasing the hazard of dementia. In addition, memory loss (LMD) had significant effect for both classes but with fairly different effect sizes. In contrast, the effects of worse executive functions (TB and DS) were statistically significant only for the second class, 
while problems in language domain (BN), functional abilities (IADLs), behaviors (NPI-Q) and age (AGE) had significant effect only for the first class. 

Combining our class probability submodel and class-specific survival submodel, we were able to correspond the two data-driven classes to meaningful clinical MCI subgroups. The first class were younger multi-domain amnestic MCI patients with early onset of language problem, which might be relevant to primary progressive aphasia occurring before memory related symptoms \citep{rogalski2016}. In contrast, the second class were older multi-domain amnestic MCI patients with impaired executive functions, which appeared to be have more typical symptoms of Alzheimer's Disease.

\subsection{Assessment of goodness-of-fit and prediction performances}

\begin{figure}
\centering
  \begin{tabular}{@{}c@{}}
    \includegraphics[width=\textwidth]{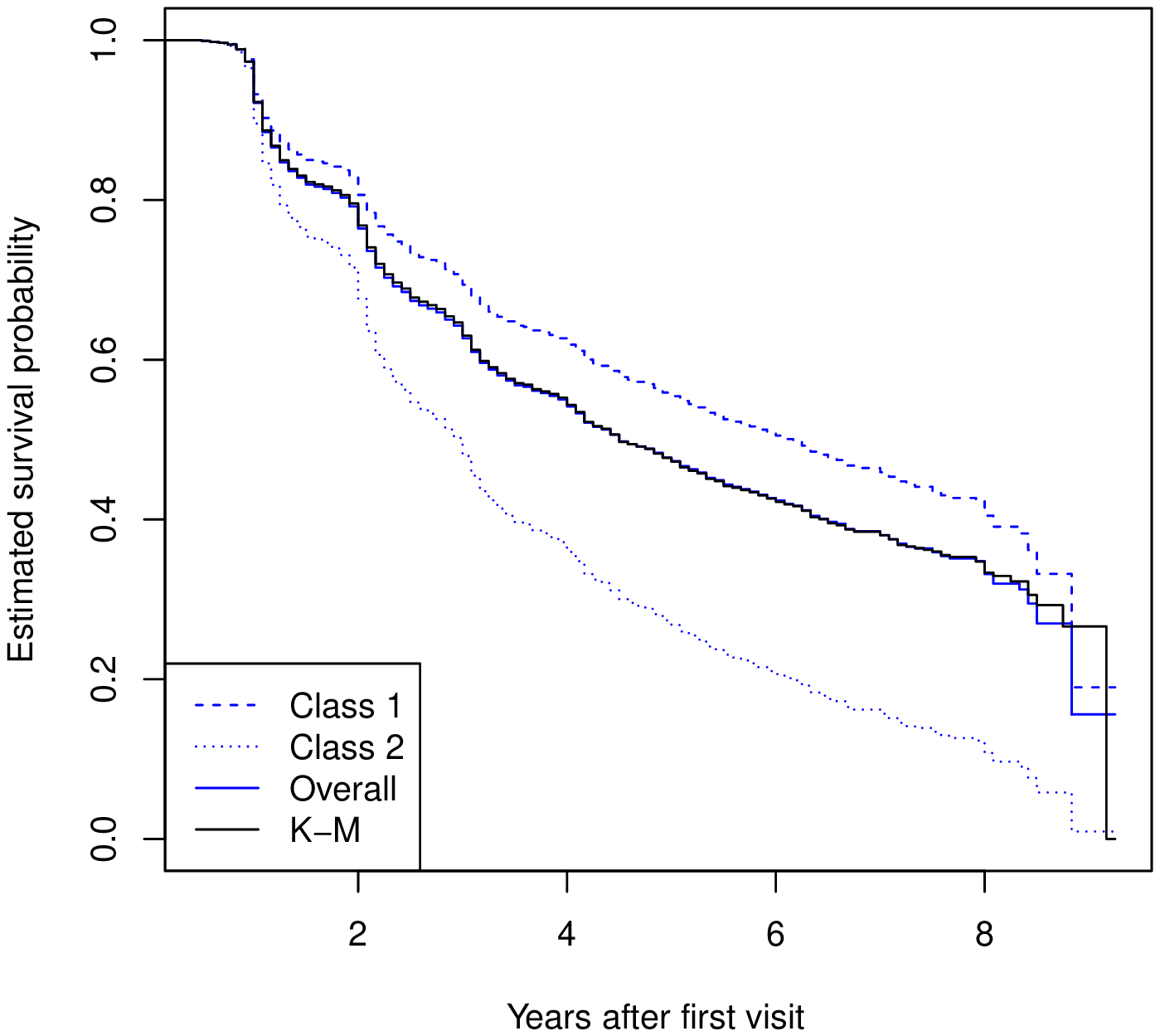}
  \end{tabular}
    \caption{Blue dashed and dotted lines (Class 1 and Class 2): Predicted class-specific survival probabilities by the latent class model. Blue solid line (Overall): Predicted overall survival probability by the latent class model. 
    K-M: Estimated Kaplan-Meier curve for overall survival probability.}
    \label{fig:real_survprob}
\end{figure}

{We assessed the goodness-of-fit of our latent class model by comparing the Kaplan-Meier curve of time-to-dementia for the MCI population, and the estimated survival probability curve from the model, calculated by averaging the predicted survival probability \eqref{eq:predicted} for all patients at each uncensored event time. As shown in Figure \ref{fig:real_survprob}, the Kaplan-Meier curve (referred to as ``K-M'') is very close to the survival curve based on the proposed model (referred to as ``Overall''), indicating reasonable goodness-of-fit. In Figure \ref{fig:real_survprob}, we also plot the average class-specific survival probabilities for all patients at each observed event time. As observed, the survival curve for class 1 is higher than the curve for class 2, indicating that patients in class 1 had slower progression towards dementia.}

\begin{figure}
\centering

  \begin{tabular}{@{}c@{}}
    \includegraphics[width=0.48\textwidth]{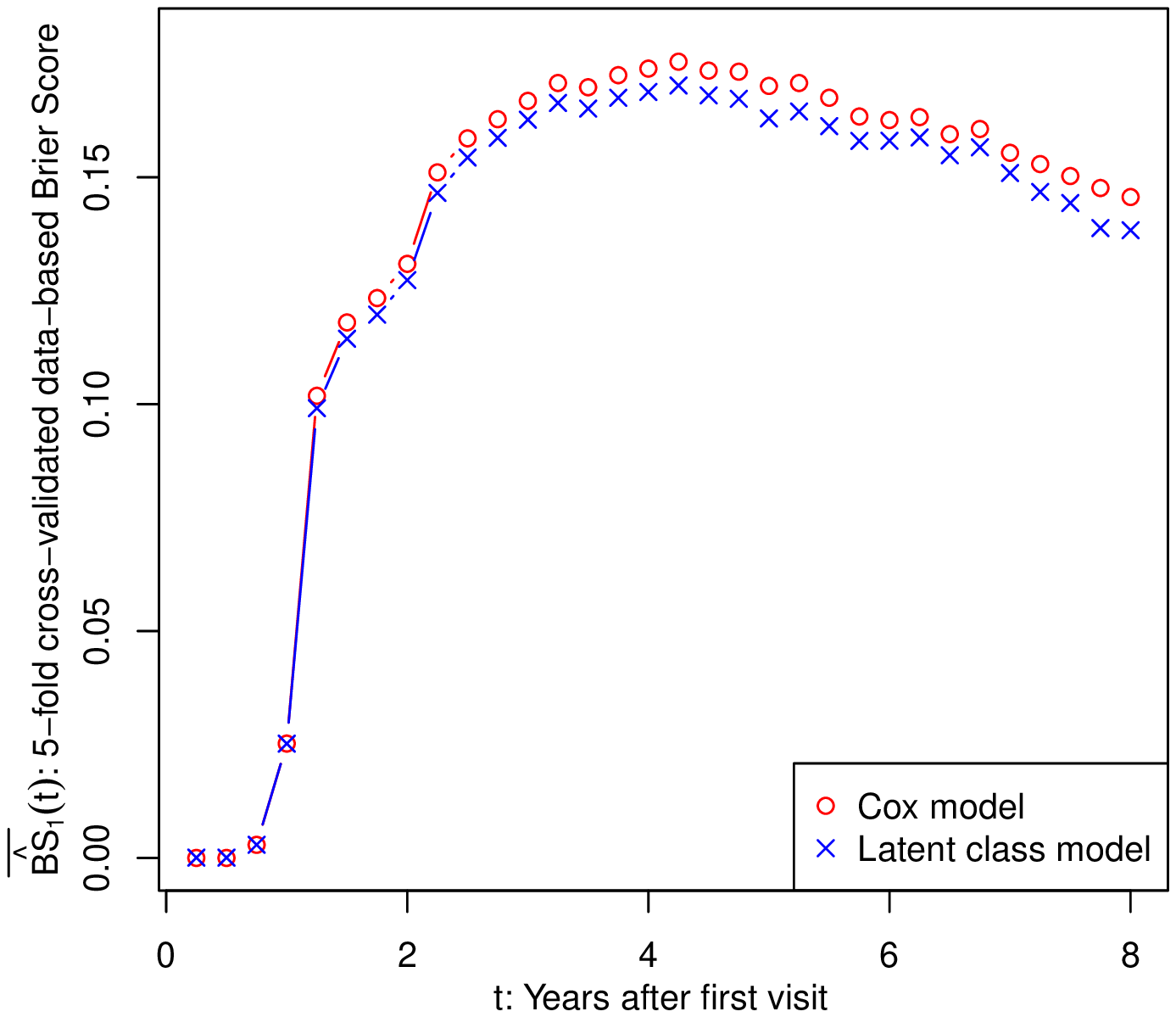}
    \includegraphics[width=0.48\textwidth]{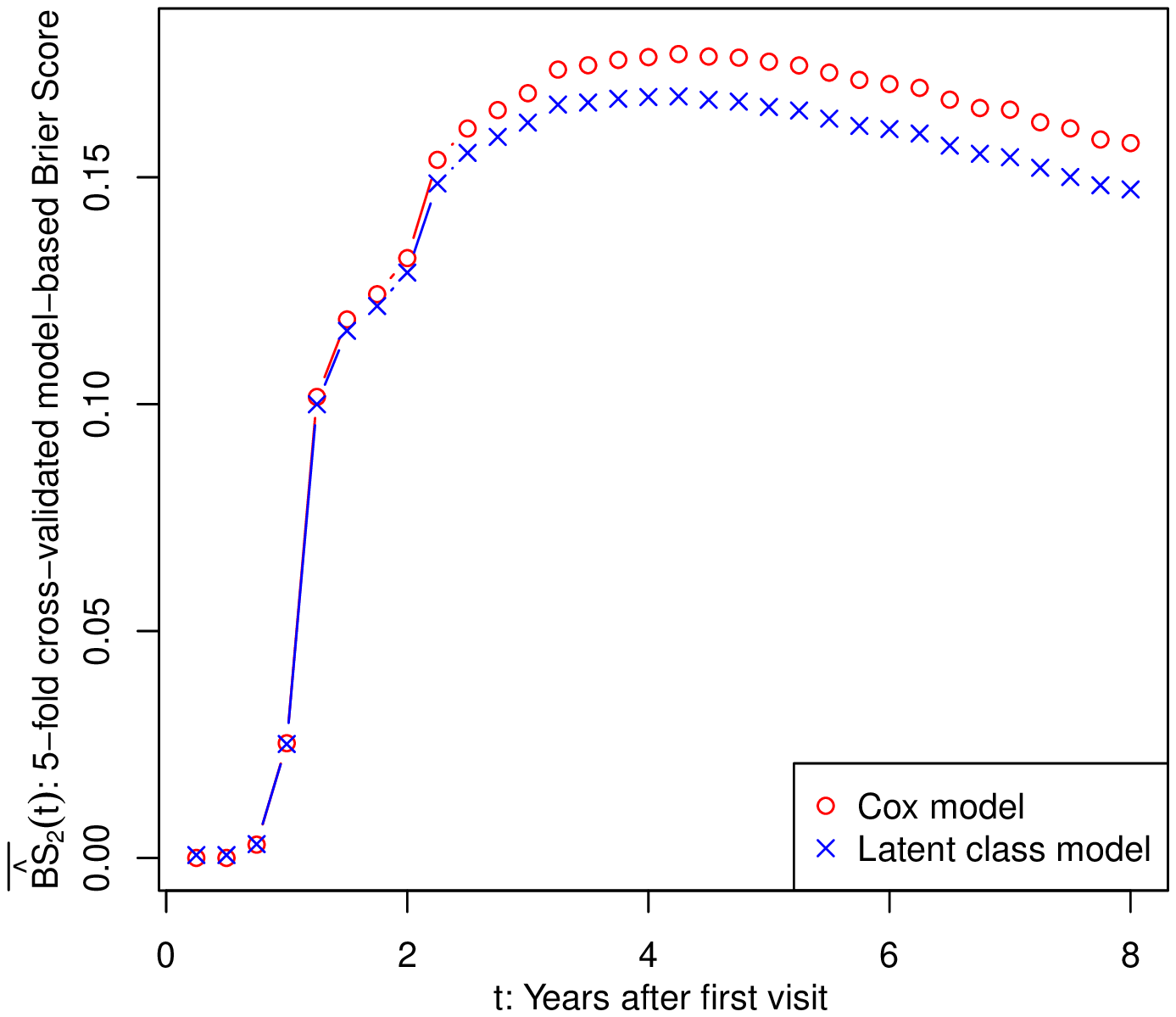} 
  \end{tabular}
\caption{Average of 5-fold cross-validated Brier Scores, $\overline{\hat{\rmn{BS}}}_j(t), j = 1,2$, obtained by the Cox model and the proposed latent class model with $L = 2$, for the UDS data application.}
    \label{fig:real_brier}
\end{figure}

{As we did in simulation, we compared the proposed method and the Cox regression model in prediction by cross-validated average Brier Scores $\overline{\hat{\rmn{BS}}}_1(t)$ and $\overline{\hat{\rmn{BS}}}_2(t)$ for $t \in (0,8]$. As shown in Figure \ref{fig:real_brier}, our proposed method achieved lower Brier Scores in five-fold cross validation, which is consistent with our observation in the simulation study. Similar to simulation scenario (IV), in real application we also observe separation of covariate effects (Table \ref{tab:realpointest}) in submodels for covariates MMSE, LMD, and NPI-Q, which explains the big improvement in Brier Scores made by the latent class model. These further demonstrated that the prediction of the survival outcome can be improved by capturing the mixture structure of a heterogeneous population.}

\section{Discussion}

In this article, we propose a semi-parametric approach to jointly modeling the latent class structure and the time-to-event outcome. By utilizing non-parametric maximum likelihood estimator (NPMLE) technique, the proposed method 
facilitates valid inference for both covariate effects and hazard functions following rigorous asymptotic theory, and is expected to be more robust than fully parametric methods. Our method also flexibly captures class-specific covariate effects in both latent class membership probabilities and class-specific hazard functions. 

Instead of including both longitudinal and time-to-event information in the joint framework, we only consider time-to-event outcome in our method. Our treatment circumvents the popular but unreliable conditional independence assumption. Based on a similar finite mixture structure as used in the proposed method, further extensions can be studied to account correlated structure of longitudinal data and survival data, while keeping the robust semi-parametric submodels developed in this method and for longitudinal observations \citep[for example]{hart20}.

Computationally, we develop a stable EM algorithm which ensures increasing observed data likelihood in each iteration. The algorithm is efficiently implemented in \texttt{Rcpp} \citep{RcppArmadillo} format and is publicly available as an R package.

\backmatter

\section*{Acknowledgements}

This work was supported by NIH grants R01 HL113548 and R01 AG055634. The authors wish to thank National Alzheimer’s Coordinating Center for making the Uniform Data Set available for our analysis.

\bibliographystyle{biom} 
\bibliography{reference}

\label{lastpage}

\end{document}